\documentclass[preprint]{iucr}              % DO NOT DELETE THIS LINE
\usepackage{graphicx}

     \journalcode{S}              % Indicate the journal to which submitted
\begin{document}                  % DO NOT DELETE THIS LINE

     %-------------------------------------------------------------------------
     % The introductory (header) part of the paper
     %-------------------------------------------------------------------------

     % The title of the paper. Use \shorttitle to indicate an abbreviated title
     % for use in running heads (you will need to uncomment it).

\title{TEMPUS, a Timepix4-based system for event-based X-rays detection}
%\shorttitle{Short Title}

     % Authors' names and addresses. Use \cauthor for the main (contact) author.
     % Use \author for all other authors. Use \aff for authors' affiliations.
     % Use lower-case letters in square brackets to link authors to their
     % affiliations; if there is only one affiliation address, remove the [a].

\cauthor[a,b]{Jonathan}{Correa}{jonathan.correa@desy.de}{address if different from \aff}
\author[c]{Alexandr}{Ignatenko}
\author[a,b]{David}{Pennicard}
\author[a,b]{Sabine}{Lange}
\author[a,b]{Sergei}{Fridman}
\author[g]{Sebastian}{Karl}
\author[a]{Leon}{Lohse}
\author[d]{Bj\"{o}rn}{Senfftleben}
\author[a]{Ilya}{Sergeev}
\author[a]{Sven}{Velten}
\author[a]{Deepak}{Prajapat}
\author[a]{Lars}{Bocklage}
\author[a,b]{Hubertus}{Bromberger}
\author[a,b]{Andrey}{Samartsev}
\author[h]{Aleksandr}{Chumakov}
\author[a,h]{Rudolf}{R\"{u}ffer}
\author[g]{Joachim}{von Zanthier}
\author[e,f,c,a]{Ralf}{R\"{o}hlsberger}
\author[a,b]{Heinz}{Graafsma}

\aff[a]{Deutsches Elektronen-Synchrotron DESY, Notkestraße 85, 22607 Hamburg, \country{Germany}}
\aff[b]{Center for Free-Electron Laser Science CFEL, Deutsches Elektronen-Synchrotron DESY, Notkestraße 85, 22607 Hamburg, \country{Germany}}
\aff[c]{Friedrich Schiller University Jena, 07743 Jena, \country{Germany}}
\aff[d]{European XFEL GmbH,	Holzkoppel 4, 22869 Schenefeld, \country{Germany}}
\aff[e]{Helmholtz-Institut Jena, 07743 Jena \country{Germany}}
\aff[f]{GSI Helmholtzzentrum für Schwerionenforschung GmbH, 64291 Darmstadt, \country{Germany}}
\aff[g]{University of Erlangen-Nuremberg, Schloßplatz 4, 91054 Erlangen, \country{Germany}}
\aff[h]{European Synchrotron Radiation Facility ESRF, 71 Av. des Martyrs, 38000 Grenoble, \country{France}}

     % Use \shortauthor to indicate an abbreviated author list for use in
     % running heads (you will need to uncomment it).

%\shortauthor{Soape, Author and Doe}

     % Use \vita if required to give biographical details (for authors of
     % invited review papers only). Uncomment it.

%\vita{Author's biography}

     % Keywords (required for Journal of Synchrotron Radiation only)
     % Use the \keyword macro for each word or phrase, e.g. 
     % \keyword{X-ray diffraction}\keyword{muscle}


     % PDB and NDB reference codes for structures referenced in the article and
     % deposited with the Protein Data Bank and Nucleic Acids Database (Acta
     % Crystallographica Section D). Repeat for each separate structure e.g
     % \PDBref[dethiobiotin synthetase]{1byi} \NDBref[d(G$_4$CGC$_4$)]{ad0002}

%\PDBref[optional name]{refcode}
%\NDBref[optional name]{refcode}

\maketitle                        % DO NOT DELETE THIS LINE

\begin{synopsis}
A full description of the TEMPUS system for photon science is given in this paper. The detector takes advantage of the new Timepix4 readout chip and, in particular, implement the use of the time-stamping mode for high resolution timing applications.
\end{synopsis}

\begin{abstract}
A readout system based on the Timepix4 ASIC, is being developed for photon science. The TEMPUS detector can be operated in two distinct modes: a photon counting mode, which allows for conventional full-frame readout at rates up to 40 kfps; and an event-driven time-stamping mode, which allows excellent time resolution in the nanosecond regime in measurements with moderate X-ray flux.
In this paper, we introduce the initial prototype, a single-chip system, and present the first results obtained at PETRA III and ESRF.
\end{abstract}

     %-------------------------------------------------------------------------
     % The main body of the paper
     %-------------------------------------------------------------------------
     % Now enter the text of the document in multiple \section's, \subsection's
     % and \subsubsection's as required.

\section{Introduction}

DESY (Deutsches Elektronen-Synchrotron) in Hamburg, Germany, operates one of the most intense storage ring light sources (SR) in the world: PETRA III [1]. The upcoming upgrade of the facility to a 4$^{th}$ generation SR, the PETRA IV project [2], will increase the brilliance by orders of magnitude, enabling for example highly-time-resolved measurements and rapid high-resolution scanning of macroscopic samples. In order to take full advantage of its elevated brilliance, and to record as many photons as possible with their relevant properties like energy, momentum, arrival time, etc., DESY is developing fast and efficient X-ray detectors [3]. Photon counting detectors are widely used at synchrotron beamlines, due to their sensitivity to single photons and high dynamic range. The LAMBDA detector system [4], based on the Medipix3 chip [5], has become the detector of choice for a variety of beamlines. The main reason for this is the combination of small pixel size, 55 $\mu$m, and a relatively high frame rate, up to 2 kfps (kilo frames per second) in 12 bit depth counter. %Moreover, the chip is two-side buttable which allows tiling up several of them to cover large areas. 
A new readout ASIC (Application-Specific Integrated Circuit) has been recently produced by the Medipix4 collaboration: Timepix4 [6], which goes beyond the photon-counting concept, by extracting more information from each photon. On the one hand, it can work in a full-frame readout mode with much higher count rate and frame rate than Medipix3. On the other hand, it can work in the time-stamping mode improving significantly over the Timepix3 chip [7]. 

The TEMPUS (Timepix4-based Edgeless Multi-Purpose Sensor) detector is being developed as a replacement to LAMBDA. In addition, as will be discussed later, the timing capabilities of the new chip open the door to other applications such as nuclear resonance scattering techniques (NRS) and and X-ray photon correlation spectroscopy (XPCS) on (sub-)$\mu$s time scales. In this paper, the system in its current state is described in Section~\ref{system}. The results of the very first experiments at PETRA III and ESRF, together with some preliminary calibration are reported in Section~\ref{experiment}. Finally, in Section~\ref{conclusions}, the project plans for the immediate future are discussed.

\section{System Description}
\label{system}

In order to boost the development speed in this first stage, and bring the system to the diverse user communities as soon as possible, a number of decisions were made during the design phase, and will be discussed now. 

\begin{figure}
	\caption{The TEMPUS single-chip prototype in its housing. The design of the carrier board allows for minimal dead area when a second system is placed on the top. Inputs to the system such as bias voltage, external trigger or digital pixels can be seen. Also, the optical cables connecting to the slow control and the high-speed data-links are visible.}
	\includegraphics[width=0.95\textwidth]{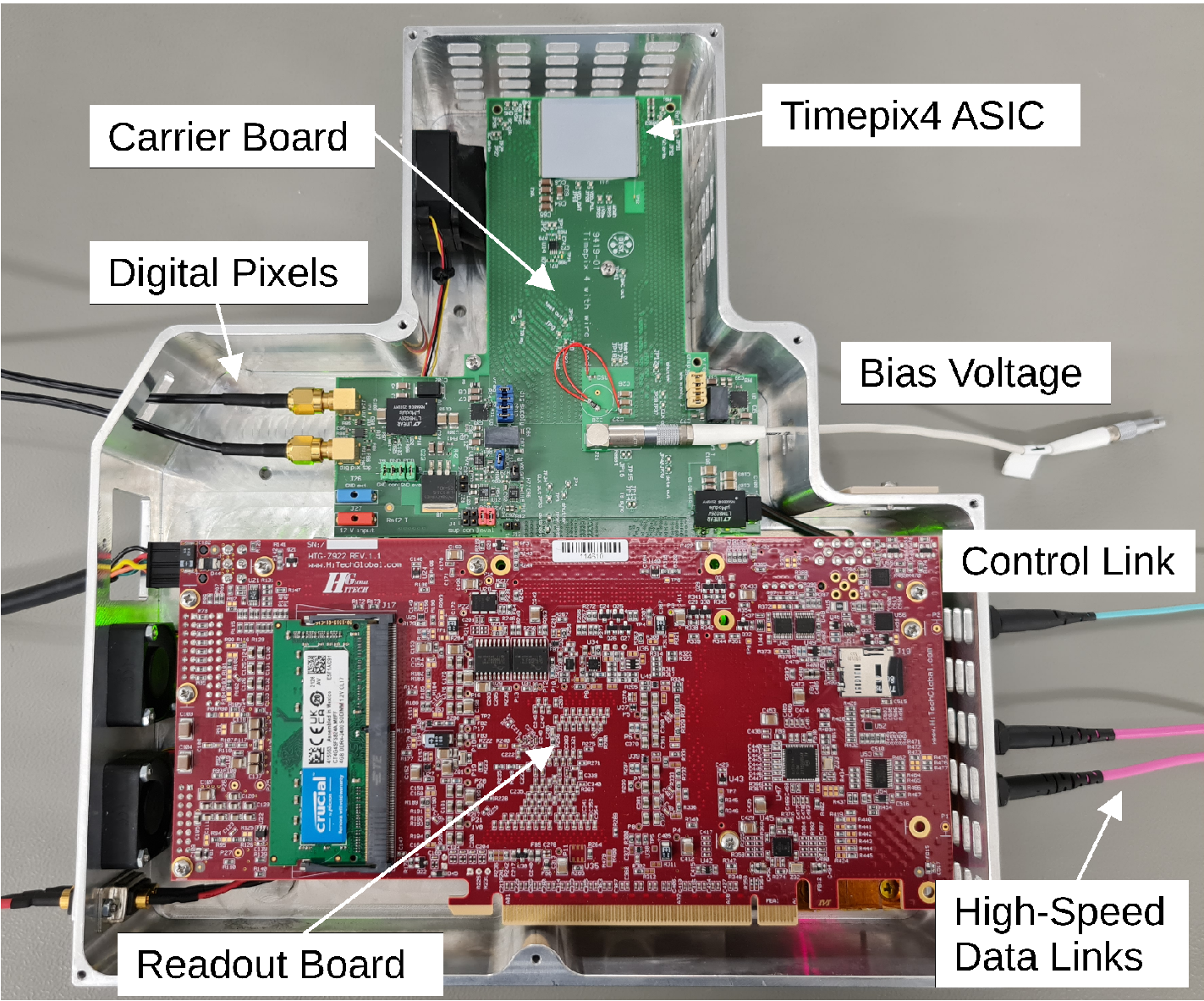}
	\label{figure_system}
\end{figure}

\begin{figure}
	\caption{Operation scheme of the TEMPUS single-chip prototype. Besides the already mentioned boards, the different inputs (triggers, etc.) and also the control and data links are shown.}
	\includegraphics[width=0.6\textwidth]{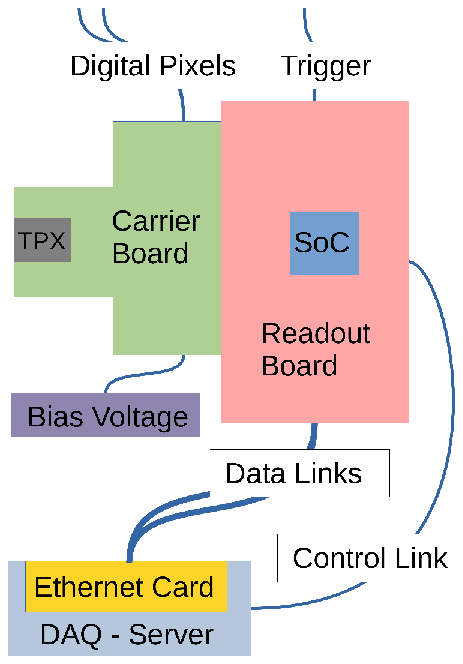}
	\label{figure_scheme}
\end{figure}

First and foremost, the larger area of the Timepix4 chip compared to previous generations (3.5 times larger than Medipix3) allows us to cover a sufficiently large solid angle without mounting several chips in this first prototype. We therefore opted for a single-chip approach. Also, little optimisation was made in terms of form factor in the design of the chip carrier board. Finally, we opted to use a powerful and commercially available Xilinx System-on-Chip (SoC) evaluation board for readout. All these different components will now be individually described in this section. A picture of the TEMPUS single-chip readout prototype is shown in Figure~\ref{figure_system}, whereas a full scheme of the system is shown in Figure~\ref{figure_scheme}.

\subsection{The Timepix4 ASIC}

The Timepix4 is a new ASIC developed by the Medipix4 collaboration aiming to improve upon both the Medipix3 and Timepix3. The chip is almost 4 times larger than its predecessor -- 448 x 512 pixel matrix -- while keeping the same pixel size -- 55 $\mu$m. The design is optimised for the use of through silicon via (TSV) technology, allowing for 4-side buttability. This will significantly reduce the dead areas between modules when multi-modules approaches for large areas are used. To allow higher readout speeds, 16 high-speed gigabit wire transmitter (GWT) links are implemented in the chip, each of which can potentially output data at readout speeds up to 10.24 Gb/s. (When discussing frame rates and event rates for TEMPUS, we assume a more conservative speed of 5.12 Gb/s per link.)
%In a nutshell, the main differences between Medipix3 and Timepix4 are described in Table \ref{table}. 
As in Medipix3, the Timepix4 chip can be operated in a photon counting mode. Hits are counted when input signal pulses in each pixel are higher than a predetermined threshold level. In this mode, the total number of hits per pixel during a frame is recorded in either 8 or 16-bit deep counters. Frame rates of up to 40 kfps in a continuous read-write mode (CRW) can be achieved. 
Much like Timepix3, Timepix4 can be also operated in a time-stamping -- or event-based -- mode. In this case, two parameters are registered per hit: every time the input signal goes above the threshold level, a timestamp with 200 ps binning is recorded at the rising edge of the pulse, this is the Time-of-Arrival (ToA); in addition, the pulse length or Time-over-Threshold (ToT) provides coarse energy information with around 1 keV resolution. A data packet with this information plus the pixel address is sent immediately out of the chip for further processing. For moderate or very low photons fluxes as encountered, for example, in NRS applications, the time-stamping mode could also help to decrease the large data volumes compared to other detector systems.  
In addition, the chip can also time stamp externally-supplied signals, to allow time measurement relative to some external event. This signal, denoted as digital pixel, is thereafter registered directly as an event in a customised number of pixels in the pixel matrix. The data packets generated this way follow the same data pipeline as other pixel data helping to minimise effects such as delays and jitters. 

\subsection{The chip carrier board}

Acting as a physical support for the chip, and to connect it to the outside world, a single-chip board was designed at DESY. It uses several double-stage voltage translators to convert the 12 V input to a very stable 1.25 V needed for the chip's operation. Although the form factor was not a strong requirement, the chip was placed at one edge of the carrier board to allow for more practical use in future experiments, for example to make it possible to surround a sample on up to 4 sides with multiple systems. 
The choice of material, Megtron6, as well as the electrical routing has been carefully done to allow the chip's operation at its full performance. In particular, special care was made to ensure that the high-speed traces from the Timepix4 GWT links to the readout board would have good signal integrity at 5.12 Gb/s.
Connectors and routing of other signals such as the sensor bias voltage, and digital pixel used for time referencing can be found on the board as well. 
An array of thermal vias has been placed in the area directly below the chip to increase the heat exchange. % However, a number of options are been considered for future implementations. This includes the use of micro-channel cooling technology integrated in a silicon interposer [9].

\subsection{The readout board}

As mentioned earlier, this first prototype was developed using a commercial Xilinx SoC evaluation board as a readout board. Specifically, the HTG-Z922: Xilinx ZYNQ ® UltraScale+ ™ MPSoC PCIe development platform was chosen.
This SoC incorporates a CPU that can run a Linux operating system, and an FPGA (Field Programmable Gate Array) fabric. For high-speed IO, this particular device offers up to 32 GTH/GTY ® transceivers which can individually deal with data rates of 16.3 and 32.75 Gb/s respectively [8]. 
During operation, the CPU is used to communicate with the control PC via a 1 Gigabit Ethernet (GbE) link using TCP. For testing purposes, a user can connect to the CPU on the Zynq directly using ssh. Python code running on the CPU is used for slow control of the Timepix4 chip, such as configuring the chip based on configuration files stored in the SoC. To do this, the CPU interfaces with the FPGA fabric, which is connected to the chip’s IOs. As well as interfacing with the chip, the FPGA executes fast control sequences, such as starting an acquisition in response to an external trigger signal. The high-speed readout is implemented using the SoC’s GTH transceivers; 16 of these receive data from the Timepix4 chip’s GWT links, and then after data aggregation in the FPGA, the data is sent to the DAQ PC over Firefly ® optic fibers with 100 GbE, using UDP as a protocol. In the future, the FPGA fabric could potentially also be used to directly implement data correction, calibration and reduction algorithms in the device, among other functionalities.

\subsection{Mechanics}
There are three major criteria concerning the prototype's housing. Firstly, it should provide protection as well as mechanical stability to the different boards mentioned above. Secondly, it should supply an easy way to be mounted to different experimental set-ups.
The third main criterion is the ability to cool all electronics components. To transfer the heat generated by the different elements, several fans have been installed in the housing to increase air circulation. The aim is to keep operation temperature stable. For reference, Timepix4 has a comparable power consumption to Medipix3, of around 1 W/cm$^{2}$.

\subsection{The Data Acquisition System (DAQ)}

At the other end of the high-speed Firefly optic fibers, we have placed the Mellanox MCX516A-CDAT Connectx-5 Ethernet card. It consists of a 100 GbE dual-port QSFP28 directly connected to the DAQ server PC with PCIe4.0. Currently during operation, the data is typically written directly to disk after it is received, and then processed offline. 
One of the main challenges ahead for this project is how to deal with the potentially very high data rates produced by the chip. A single chip working at a fast frame rate could already produce up to 80 Gb/s of data. These numbers will linearly scale up when multi-chip systems will be produced to cover large active areas. It is foreseen that an accelerator card with built-in network links - such as a Xilinx Alveo FPGA card or Nvidia A100X GPU card - could be used instead in the future for further data processing (including data reduction) prior to saving it in to disk. High-speed data acquisition using accelerator cards has been demonstrated for other detectors; see for example [9].

\subsection{Control and data processing software}

As mentioned above, Python code running on the CPU in the SoC provides various functions that control the Timepix4 chip and the detector as a whole – for example, configuring the detector and starting data taking. In turn, higher-level control software can call on these functions to do more sophisticated acquisition sequences. Currently, this also generally consists of running Python scripts directly on the SoC, by connecting to it via ssh. In the future, a gRPC interface will be implemented to allow beamline control systems and other software to issue commands to the detector. Data reception on the PC is currently done using Python code to receive data packets from the 100 GbE link and save them to disk; the raw data stream from the detector can then be converted into meaningful data offline.

%In this first iteration, a number of Python scripts are used to operate the Timepix4 chip. As mentioned above, the control scripts are kept in and run from the SoC, whereas for the data readout, this is done at the DAQ server. For the latter, it is foreseen to use much more performing C++ scripts at a later stage.

%Before operating a Timepix4 chip, a number of steps are needed. On the one hand, a set of DAC (Digital to Analog Converters) settings need to be fine tuned depending on factors such as the operation mode, incoming flux, etc. Also, a threshold equalisation is performed in order to locally adjust the pixel threshold, accounting for variations of the pixel's baselines, and therefore, have a more uniform response across the detector. This algorithm also results in the creation of pixel masks where hot or noisy pixels are hidden out. 

Some work has been done at DESY to readout Timepix3-based systems in the past. In particular, the Controlled Molecule Imaging (CFEL/CMI) group, has developed a control, visualisation, and acquisition software called Pymepix to run commercially available Timepix3 devices [10]. Within this context, work has already started on a next version of this software to be fully compatible with the new Timepix4 chip. At this point in time, this package already allows us to have a real time online visualisation of the data generated by TEMPUS. It is also foreseen for this software to eventually fully control the detector system. Some or all of this software may be upgraded to a faster language like C++ to improve throughput; Pymepix is structured so that performance-critical code blocks can be replaced without having to rewrite the entire software in the new language.

Additionally, software has been developed to calibrate and configure new Timepix4 chips before use.  Firstly, the Timepix4 chip contains a set of Digital to Analog Converters (DACs) which provide various reference voltages and currents; these settings need to be fine tuned depending on factors such as the operation mode, incoming flux, etc. Also, each pixel has adjustable circuitry that can compensate for pixel-to-pixel variations in threshold to improve the image uniformity, and a threshold equalisation is performed in order to find the optimal setting for each pixel. This algorithm also leads to the creation of pixel masks where hot or noisy pixels are electronically masked so they don’t produce spurious data packets. 
%The Timepix4 chip offers the possibility to readout data via the same slow link (1Gb/s) used for device control. This debugging functionality has proof very useful for testing the readout system. Also, and albeit reducing the maximum hit rate of our system to below 5000 hits per second, it has allowed for the preliminary use of the system as it will be shown in the next Section.

\section{Testing the system}
\label{experiment}

The first tests in the lab were performed with a very restricted data bandwidth. In particular, a debugging feature of the chip was used to do the readout through the slow control link that is normally used to configure and control the chip, rather than the high-speed links. The hit rate capability of the system was therefore limited to roughly 5000 hits per second. This was however enough for testing basic functionalities, calibration, and pixel equalisation. %
Regardless of this limitation in the event rate, the system could also be used in experiments where the photon flux is low and high time resolution is a requirement. Therefore, we aimed for users/beamlines, which may be interested in the time-stamping mode, have some background in fast (e.g. nanosecond) time resolved experiments and the necessary infrastructure. The two NRS beamlines, at PETRA III and at ESRF, P01 and ID14 respectively, fit those demands. Further, they provide so-called high resolution monochomators (HRM) with meV energy resolution at 14.4 keV, which allow for a very clean x-ray beam without any traces of harmonics etc.. These tests helped to speed up the development: from the fine tuning of the control software and the preliminary data analysis tools, to the improvement of our understanding of the chip and other components of the system.

A TEMPUS detector using a 300~$\mu$m-thick p-on-n silicon sensor produced by Advafab OY [10] was tested at the PETRA III - P01 beamline. When detecting X-ray photons, the main factor limiting the time resolution is that photons can be absorbed at varying depths in the sensor, resulting in varying drift times to the pixel contacts. This effect will depend on photon energy; at higher photon energies the longer absorption length will lead to greater variation in the absorption depth. To achieve the best time resolution, ideally an electron-collecting sensor design should be used, to take advantage of electrons’ much higher mobility than holes, and the sensor should be operated at a high voltage. Unfortunately, only a hole-collecting sensor was available, and high leakage currents were seen when biasing the sensor so we limited the voltage to around 100 V. Under these sub-optimal conditions, we estimate a value of around 20 ns timing resolution for the photon energy used during this experiment. The description and results of these tests are found in Section~\ref{experiment_petra}

Only a few months after the initial tests at P01, the system was tested at the nuclear scattering beamline ID14 at ESRF. In this case a single high-speed data link running at 1.28 Gb/s was used which pushed the maximum event-rate to around $2\times10^{7}$ hits per second. Also, a similar 300~$\mu$m-thick p-on-n sensor was used, allowing for a higher bias voltage (up to 200 V), which increased the carrier drift velocity and thus improved the time resolution down to a few ns. Results of this experiment are presented in Section~\ref{experiment_esrf}

%In this section we describe both experiments and discuss the results obtained. 
%We Also, two different 300~$\mu$m-thick p-on-n silicon sensor produced by Advafab OY [12] were used in this first test. Unfortunately, high leakage currents were seen when biasing the sensor so we limited voltage to around 100 V. Under these conditions, a worsening of the timing resolution of the system due to a slower electron/holes drift velocity within the sensor is expected. We estimate a value of around 20 ns timing resolution for the photons used during this experiment. 

\subsection{Tests at the P01 nuclear resonance beamline}
\label{experiment_petra}

The P01 beamline at PETRA III [12] is dedicated to NRS and Inelastic X-ray Scattering (IXS and RIXS) experiments at photon energies between 2.5 keV and 90 keV. The beamline offers high energy resolution monochromators in the meV regime. The maximum beam size is in the order of 1mm $\times$ 1mm. For the measurements, instead impinging this beam directly on the much bigger sensor, we used the X-rays which were scattered from various target samples under $90^{\circ}$ in a horizontal scattering geometry. This allowed us to illuminate the entire surface of the detector more or less homogeneously. Although we used several metals as targets to record elastically scattered photons and characteristic fluorescence radiation, particular focus was put on $^{57}$Fe. For this purpose, the photon energy was set to 14.4 keV. The SR was running the so-called \textit{timing mode} during the tests [13]. This mode consists in a total of 40 electron bunches per revolution. A full revolution takes 7.685 $\mu$s and the bunches are equally spaced at around 192 ns.

Note that the beam energy was set about 30 meV above resonance so that the 14.4 keV nuclear level was excited via inelastic process where phonons in the sample were created via energy transfer of the photons to the crystal lattice. This off-resonance excitation of the nuclear resonance ensured a fully incoherent emission of the nuclear fluorescence photons. Further studies on the nuclear transition of $^{57}$Fe performed with the TEMPUS detector will follow this work.

\begin{figure}
   	\caption{(a), ToT vs. ToA plot shows the phenomenon known as time-walk. Signals with lower amplitude are registered at a later ToA due to their later crossing of the threshold. (b), The ToT vs. ToA correlation is now shown after the time-walk correction is applied. Lower ToT hits are now registered at the same time than the higher ones.}\includegraphics[width=0.95\textwidth]{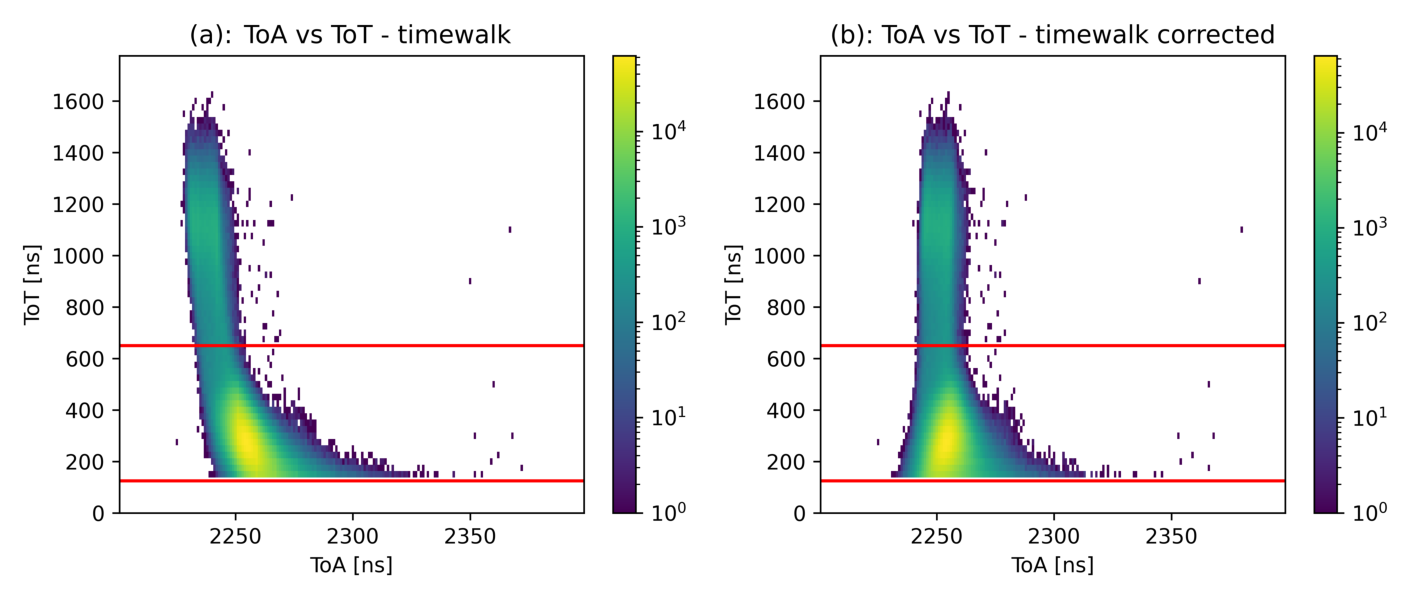}
   	\label{time_walk}
\end{figure}

% An image of one of the targets used can be seen Figure~\ref{figure_system}~B. Under this configuration, two types of photons will be reaching the sensor, the scattered ones and the ones produced by fluorescence. 
The detector was placed approximately orthogonal to the incoming beam and at an distance of $\sim$ 5 cm from the target. In this configuration, essentially elastically scattered photons, which have the same energy as the primary beam, and fluorescence photons, which are at a lower energy which depends on the target element, can be observed. Timepix4 can simultaneously measure energy and time information from individual events. Under certain conditions, this enables the separation of the different photon contributions (e.g. background, scattering, fluorescence) arriving to the detector. In the retrieval of the timing information, the so-called time-walk effect plays particularly a role for low amplitude signals: simultaneous signal pulses of different amplitude discriminated by a constant threshold are measured at different times. For the low amplitude signals it takes significantly longer to reach the threshold than for larger one. In consequence, the correlation of ToT and ToA for simultaneous events, reflects the time-walk effect as shown in Figure~\ref{time_walk}~(a). As a result a function can be fitted to the correlation and for each event a ToA correction can be derived from its ToT measurement. The outcome of the correction is displayed in Figure~\ref{time_walk}~(b). Note that for events with very low ToT, already close to the threshold value, the implemented time-walk correction shifts the ToA towards lower values. In the extreme case, some of them get corrected ToA values lower than the prompt photons. This is an artifact of the time-walk correction itself which is done to improve the time resolution only. This effect has already been studied for Timepix3 [14].

\begin{figure}
	\caption{ToT spectrum; after setting a threshold at 100 ns ToT, two contributions: the first one at around 300 ns ToT, which we attribute mainly to the 6.4 keV K-$\alpha$ fluorescence energy of Fe, and a second one at around 1100 ns ToT, which we attribute to the scattered photons at 14.4 keV. We set an arbitrary cut-off energy at 600 ns ToT and split all events into these two distributions. 
	}\includegraphics[width=0.80\textwidth]{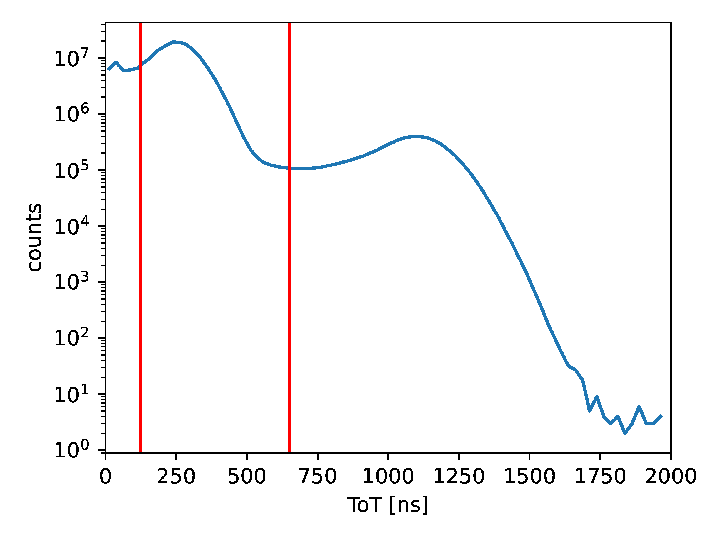}
	\label{ToT_1D}
\end{figure}

The distinction of different photon energies can be demonstrated by studying the ToT distribution of the data shown in Figure~\ref{ToT_1D}. For single photon events the signal amplitude and in consequent ToT scales with the photon energy. After setting a threshold at 100 ns ToT, the first maximum is found at around 300 ns ToT, which we attribute mainly to the 6.4 keV K-$\alpha$ fluorescence energy of Fe. We found the second maximum at around 1100 ns ToT, which we attribute to the scattered photons from $^{57}$Fe at 14.4 keV. Note that, in order to improve the energy resolution (limited in principle by the chip to 1 keV), a full per pixel energy calibration should be used. Also, clustering algorithms are usually applied to bring together single events spread onto several pixels. These analysis methods however exceed the scope of the current work. We set an arbitrary cut-off energy at 650 ns ToT and split all events into these two distributions: low and high energies. The result is that some 14.4 keV photons will be incorrectly attributed to low energy fluorescence events due to charge sharing. In particular, they will appear as a long tail of the lower energy distribution.   

\begin{figure}
	\caption{(a), events binned by ToA obtained for several revolutions of PETRA III (initialised once per revolution). Hits corresponding to the 40 different electron bunches can be clearly distinguished due to the high time resolution of TEMPUS. (b), the ToA binning of the two different TOT contributions divided by an arbitrary threshold placed at 650 ns ToT are shown. The obtained time resolution, defined as the FWHM of the distribution correspond to the expected values. The higher time resolution of the lower energy photons is due to the shorter absorption length in silicon. As mentioned in the text, time resolution was limited by the low biased silicon sensor: 23.3 and 9.7 ns respectively.}
	\includegraphics[width=0.95\textwidth]{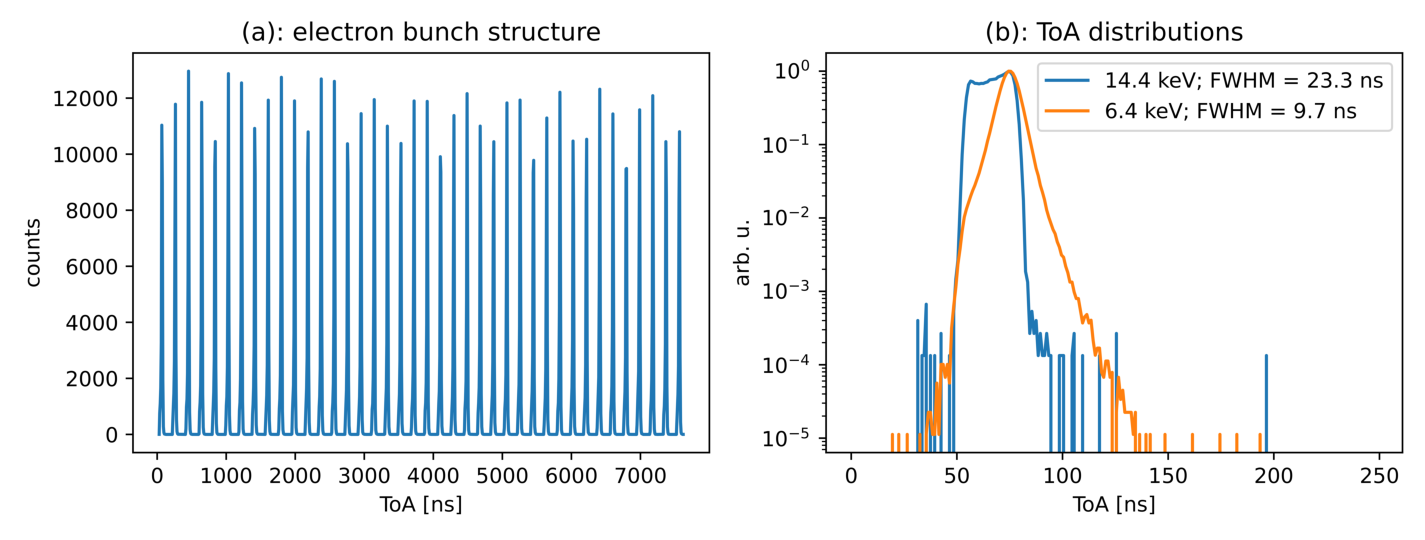}
	\label{figure_petra}
\end{figure}

We used this information to bin the events by ToA. This can be seen in Figure~\ref{figure_petra}~(a). Using an attenuated flux, we identify photons coming from each individual electron bunches produced by PETRA III. Results show the ToA of individual events with respect to the beginning of each revolution. Due to the already mentioned limitations on hit rate set by the readout system, the experiments had a relatively low duty cycle, where hits were recorded over a short time period (between 80 and 1600 $\mu$s) followed by a longer pause for readout. Therefore, data from many revolutions were combined to attain sufficient statistics. The individual bunches of the PETRA III machine operating in the timing mode are clearly visible. As mentioned above, the bunches are equally spaced 192 ns apart. Note that, for a detector working in a frame-based mode, the needed frame rate to achieve this result must be at least 5.2 MHz, with the subsequent large volume of data produced.
Due to precise timing references by the digital pixel inputs, the ToA of the events can be aligned to the timing of the bunch clock instead of the revolution clock. The so produced ToA distribution is shown in Figure \ref{figure_petra}~(b). The time resolution, in this case defined as the full width half maximum (FWHM), of 23.3 ns for high and 9.7 ns for the low energy photons is in good agreement with the expected value 20 for the high energy photons and the relatively low bias voltage used across the sensor. The very different shapes of the two distributions are also expected. The non-flat and asymmetric top of the higher energy distribution is explained by due to photons converted very close to the entrance window of the sensor. The carriers produced there, will take around 20 ns to arrive to the other end of the sensor. As for the photons converted deeper in the sensor, the travel time of the carrier is shorter, which implies a lower ToA, but due to the attenuation produced by the silicon sensor itself, their number will be smaller. The better time resolution obtained for low energy photons is explained by the fact that they are all absorbed in a very shallow thickness of the sensor.

\subsection{Tests at the nuclear scattering ID14 beamline}
\label{experiment_esrf}

The new nuclear scattering ID14 beamline at ESRF is also dedicated to NRS experiments. Here too, several metals were used as target and again emphasis was put on $^{57}$Fe. The photon energy was set to 14.4 keV and the SR was running the \textit{7/8 + 1} filling mode during the tests [15]: 7/8 of the ring is filled with 868 uniform bunches and the remaining 1/8 is filled in its centre with a relatively large single bunch. As mentioned above, a new assembly was used with a sensor that did not show any leakage current issues at higher voltage. %This mode consists in a total of 40 electron bunches per revolution. A full revolution takes 7.85 $\mu$s and the bunches are equally spaced at around 192 $n$s.

\begin{figure}
	\caption{(a), events binned by ToA obtained for several revolutions of ESRF (initialised once per revolution). The 7/8 + 1 electron bunch structure is visible. (b), the low an high energy contributions are shown as a function of ToA. Due to the higher bias voltage, an improve time resolution was achieved: 12.1 and 8.5 ns respectively.}
	\includegraphics[width=0.95\textwidth]{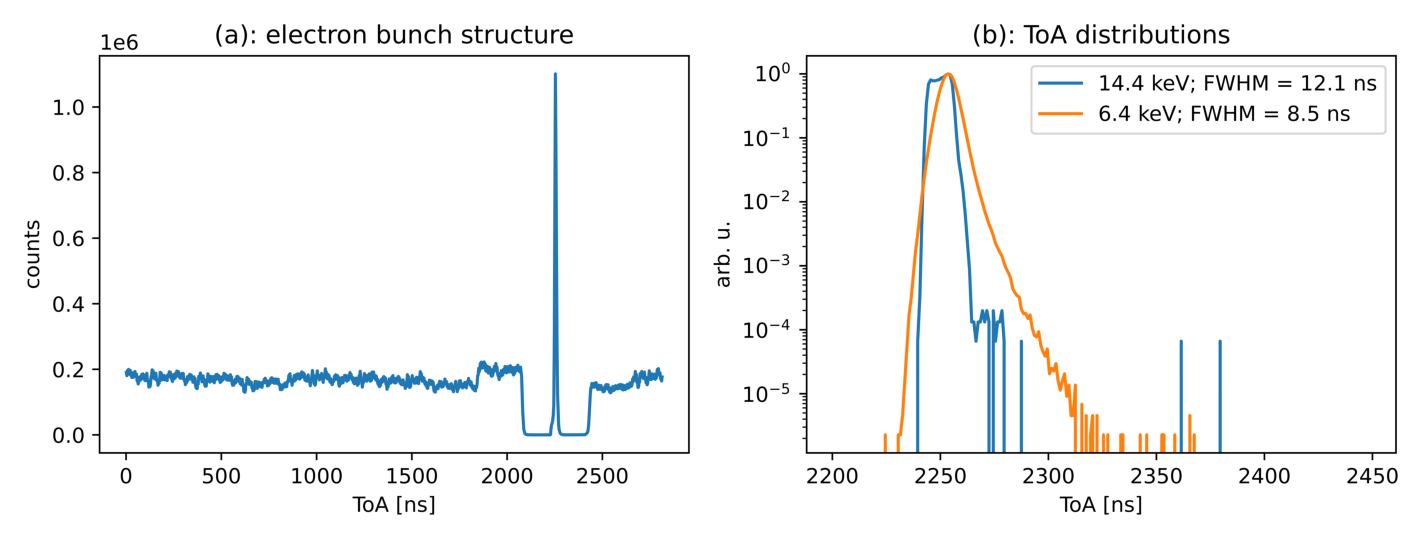}
	\label{figure_esrf}
\end{figure}

Again, after performing the time-walk correction and energy discrimination that was mentioned above, and shown in Figure~\ref{time_walk}, the ToA distribution obtained for the whole revolution is also shown in Figure \ref{figure_esrf}~(a). The structure of the electron bunches in this mode with the single-bunch in the middle is clearly visible. The ToA distribution obtained for the single-bunch seen above is also shown in Figure \ref{figure_esrf}~(b). In this case, and since the bias voltage has been increased, the time resolution achieved for the two different photon energies are 12.1 and 8.5 ns respectively. As mentioned earlier, the artifact on the time-walk correction wrongly assigned lower ToA values to events with low ToT values. That explains the fact that there seem to be events of the lower energy distribution arriving earlier than the ones from the higher energy distribution. Also, the 14.4 keV photons incorrectly attributed to the low energy distribution, due to charge sharing, explain the long tails of the latter. 
%Another experiment performed was the energy distribution of a number of different metal targets. This can be seen in Figure~\ref{figure_esrf}. 

\section{Conclusions and Outlook}
\label{conclusions}

The single-chip TEMPUS prototype is being developed by DESY with the aim to replace LAMBDA detectors currently deployed at PETRA III. The use of a new, more performing chip, -- Timepix4 -- will make possible to work at a high frame rate. In addition, it features a time-stamping mode, which provides event-based X-ray detection with few-ns time resolution. The system has been described and the prototype has seen its first light. Further characterisation and calibration tests are ongoing. Preliminary results from NRS beamline at PETRA III and ESRF show that the achievable time resolution of a few ns is currently limited by the sensor used, a 300-$\mu$m thick hole-collecting silicon sensor. This resolution is sufficient to fully resolve the 40 electron bunches of the \textit{timing mode} at PETRA III and the single-bunch in the \textit{7/8 + 1} mode at ESRF. Improvements in this area could be achieved by using different type of sensors such as electron-collecting silicon sensors, or the use of Low Gain Avalanche Detector (LGADs) technology [16], for faster signal collection. Alternatively, the spread in absorption depth can be reduced by using high-Z materials such as GaAs and CdTe. 
Moreover, data acquisition through the newly developed high-speed data-links could be demonstrated. Ongoing development will enable the use of all 16 links, and achieve their maximum data rate of 5.12 Gb/s (or total of 80 Gb/s).
Further tests with the single-chip prototype in collaboration with users are foreseen. Besides the ongoing collaboration in the field of NRS, the XPCS community is interested in this development to access the fast dynamics in the sub-$\mu$s regime. Moreover, and in collaboration with other partners, a multi-chip module is also under development.

     % Appendices appear after the main body of the text. They are prefixed by
     % a single \appendix declaration, and are then structured just like the
     % body text.

%\appendix
%\section{Appendix title}
%
%Text text text text text text text text text text text text text text
%text text text text text text text.
%
%\subsection{Title}
%
%Text text text text text text text text text text text text text text
%text text text text text text text.
%
%\subsubsection{Title}
%
%Text text text text text text text text text text text text text text
%text text text text text text text.

     %-------------------------------------------------------------------------
     % The back matter of the paper - acknowledgements and references
     %-------------------------------------------------------------------------

     % Acknowledgements come after the appendices

\ack{Acknowledgements}

The authors acknowledge funding from QuCoLiMa (Quantum Cooperativity of Light and Matter) CRC-TR 306 Project A03, as well as from the DataX project of the Innovation Pool in the Research Field 'Matter' of the German Helmholtz Association.

     % References are at the end of the document, between \begin{references}
     % and \end{references} tags. Each reference is in a \reference entry.

     %-------------------------------------------------------------------------
     % TABLES AND FIGURES SHOULD BE INSERTED AFTER THE MAIN BODY OF THE TEXT
     %-------------------------------------------------------------------------

     % Simple s should use the tabular environment according to this
     % model

\end{document}